\documentclass[runningheads]{llncs}

\usepackage{amsmath,amssymb}
\usepackage[utf8]{inputenc}
\usepackage[english]{babel}
\usepackage{amssymb}
\usepackage{mathtools}
\usepackage{listings}
\usepackage{comment}
\usepackage{indentfirst}
\usepackage{hyperref}
\usepackage{stmaryrd}
\usepackage{eufrak}
\usepackage{lstcoq}
\usepackage{placeins}
\usepackage[dvipsnames]{xcolor}
\usepackage{tcolorbox}
\usepackage{soul}
\usepackage{enumitem}
\setlist{nosep}

\let\llncssubparagraph\subparagraph
\let\subparagraph\paragraph
\usepackage{titlesec}
\let\subparagraph\llncssubparagraph

\titlespacing*{\section}{0pt}{1ex plus 1ex minus .2ex}{1ex plus .2ex}
\titlespacing*{\subsection}{0pt}{1ex plus 1ex minus .2ex}{1ex plus .2ex}
\titlespacing*{\paragraph}{0pt}{0ex}{1em}

\lstdefinelanguage{minikanren}{
basicstyle=\normalsize,
keywords={fresh},
sensitive=true,
commentstyle=\itshape\ttfamily, 
keywordstyle=\textbf,
identifierstyle=\ttfamily,
basewidth={0.5em,0.5em},
columns=fixed,
fontadjust=true,
literate={fun}{{$\lambda\;\;$}}1 {->}{{$\to$}}3 {===}{{$\,\equiv\,$}}1 {=/=}{{$\not\equiv$}}1 {|>}{{$\triangleright$}}3 {/\\}{{$\wedge$}}2 {\\/}{{$\vee$}}2,
morecomment=[s]{(*}{*)}
}

\usepackage{letltxmacro}
\newcommand*{\SavedLstInline}{}
\LetLtxMacro\SavedLstInline\lstinline
\DeclareRobustCommand*{\lstinline}{%
  \ifmmode
    \let\SavedBGroup\bgroup
    \def\bgroup{%
      \let\bgroup\SavedBGroup
      \hbox\bgroup
    }%
  \fi
  \SavedLstInline
}

\usepackage{todonotes}

\newcommand{\inbr}[1]{\left<{#1}\right>}

\newcommand{\ruleno}[1]{\mbox{[\textsc{#1}]}}

\newcommand{\sembr}[1]{\llbracket{#1}\rrbracket}

\newcommand{\dom}[1]{\mathtt{dom}\;{#1}}

\renewcommand{\dom}[1]{\mathcal{D}om\,({#1})}
\newcommand{\ran}[1]{\mathcal{VR}an\,({#1})}
\newcommand{\fv}[1]{\mathcal{FV}\,({#1})}
\newcommand{\tr}[1]{\mathcal{T}r_{#1}}
\newcommand{\step}{\circ}

\newcommand{\bigslant}[2]{{\raisebox{.2em}{$#1$}\left/\raisebox{-.2em}{$#2$}\right.}}

\let\emptyset\varnothing

\DeclareUnicodeCharacter{2212}{-}

\title{Certified Semantics for Relational Programming\thanks{The reported study was funded by RFBR, project number 18-01-00380.}}

\author{Dmitry Rozplokhas\inst{1,3} \and 
Andrey Vyatkin\inst{2} \and
Dmitry Boulytchev\inst{2,3} } 
\authorrunning{D. Rozplokhas et al.}
\institute{Higher School of Economics, Russia \and
Saint Petersburg State University, Russia \and
JetBrains Research, Russia}

\begin{document}

\setlength{\belowcaptionskip}{-5pt}
\setlength{\abovecaptionskip}{0pt}

\setlength{\abovedisplayskip}{-3pt}
\setlength{\belowdisplayskip}{-2pt}
\setlength{\abovedisplayshortskip}{0pt}
\setlength{\belowdisplayshortskip}{2pt}

\maketitle
\setcounter{footnote}{0}

\begin{abstract}
  We present a formal study of semantics for the relational programming language \textsc{miniKanren}. First,
  we formulate a denotational semantics which corresponds to the minimal Herbrand model for definite logic
  programs. Second, we present operational semantics which models interleaving, the distinctive feature of \textsc{miniKanren}
  implementation, and prove its soundness and completeness w.r.t. the denotational semantics.
  Our development is supported by a \textsc{Coq} specification, from which a reference interpreter can be
  extracted. We also derive from our main result a certified semantics (and a reference interpreter) for SLD resolution
  with cut and prove its soundness.
\end{abstract}

\section{Introduction}

In the context of this paper, we understand ``relational programming'' as a puristic form of logic programming with all extra-logical
features banned. Specifically, we use \textsc{miniKanren} as an exemplary language; \textsc{miniKanren} can be seen as
a logical language with explicit connectives, existentials and unification, and is mutually convertible to the pure logical subset of
\textsc{Prolog}.\footnote{A detailed \textsc{Prolog}-to-\textsc{miniKanren} comparison can be found here: \url{http://minikanren.org/minikanren-and-prolog.html}}
Unlike \textsc{Prolog}, which relies on SLD-resolution, most \textsc{miniKanren} implementations use a monadic \emph{interleaving
search}, which is known to be complete~\cite{SmallEmbedding}.
\textsc{miniKanren} is designed as a shallow DSL which may help to equip the host language with logical reasoning features. This
design choice has been proven proven to be applicable in practice, and there are more than 100 implementations for almost 50 languages.

Although there already were attempts to define a formal semantics for \textsc{miniKanren}, none of them were capable of reflecting the distinctive property of \textsc{miniKanren}'s search~--- \emph{interleaving}~\cite{Search}.
Since this distinctive search strategy is essential for the specification of the language and its extensions, the description of almost all development on miniKanren was not based on formal semantics.
The introductory book on \textsc{miniKanren}~\cite{TRS} describes the language by means of an evolving set of examples. In a
series of follow-up papers~\cite{MicroKanren,CKanren,CKanren1,AlphaKanren,2016,Guided} various extensions of the language were presented with
their semantics explained in terms of a \textsc{Scheme} implementation. We argue that this style of semantic definition is
fragile and not self-sufficient since it relies on concrete implementation languages' semantics and therefore is not stable under the host language replacement.
In addition, the justification of important properties of the language and specific relational programs becomes cumbersome.

In this paper, we present a formal semantics for core \textsc{miniKanren} and prove some of its basic properties. First,
we define denotational semantics similar to the least Herbrand model for definite logic programs~\cite{LHM}; then
we describe operational semantics with interleaving in terms of a labeled transition system. Finally, we prove soundness and
completeness of the operational semantics w.r.t the denotational one. We support our development with a formal specification
using the \textsc{Coq} proof assistant~\cite{Coq}, thus outsourcing the burden of proof checking to the automatic tool and
deriving a certified reference interpreter via the extraction mechanism. As a rather straightforward extension of our
main result, we also provide a certified operational semantics (and a reference interpreter) for SLD resolution with cut, a new result
to our knowledge; while this step brings us out of purely relational domain, it still can be interesting on its own.

\begin{figure*}[t]
\[
\begin{array}{cccll}
  &\mathcal{C} & = & \{C_i^{k_i}\} & \mbox{constructors with arities} \\
  &\mathcal{T}_X & = & X \cup \{C_i^{k_i} (t_1, \dots, t_{k_i}) \mid t_j\in\mathcal{T}_X\} & \mbox{terms over the set of variables $X$} \\
  &\mathcal{D} & = & \mathcal{T}_\emptyset & \mbox{ground terms}\\
  &\mathcal{X} & = & \{ x, y, z, \dots \} & \mbox{syntactic variables} \\
  &\mathcal{A} & = & \{ \alpha, \beta, \gamma, \dots \} & \mbox{semantic variables} \\
  &\mathcal{R} & = & \{ R_i^{k_i}\} &\mbox{relational symbols with arities} \\
  &\mathcal{G} & = & \mathcal{T_X}\equiv\mathcal{T_X}   &  \mbox{unification} \\
  &            &   & \mathcal{G}\wedge\mathcal{G}     & \mbox{conjunction} \\
  &            &   & \mathcal{G}\vee\mathcal{G}       &\mbox{disjunction} \\
  &            &   & \mbox{\lstinline|fresh|}\;\mathcal{X}\;.\;\mathcal{G} & \mbox{fresh variable introduction} \\
  &            &   & R_i^{k_i} (t_1,\dots,t_{k_i}),\;t_j\in\mathcal{T_X} & \mbox{relational symbol invocation} \\
  &\mathcal{S} & = & \{R_i^{k_i} = \lambda\;x_1^i\dots x_{k_i}^i\,.\, g_i;\}\; g & \mbox{specification}
\end{array}
\]
\caption{The syntax of the source language}
\label{syntax}
\end{figure*}

\section{The Language}
\label{language}
 
In this section, we introduce the syntax of the language we use throughout the paper, describe the informal semantics, and give some examples.

The syntax of the language is shown in Fig.~\ref{syntax}. First, we fix a set of constructors $\mathcal{C}$ with known arities and consider
a set of terms $\mathcal{T}_X$ with constructors as functional symbols and variables from $X$. We parameterize this set with an alphabet of
variables since in the semantic description we will need \emph{two} kinds of variables. The first kind, \emph{syntactic} variables, is denoted
by $\mathcal{X}$. The second kind, \emph{semantic} or \emph{logic} variables, is denoted by $\mathcal{A}$.
We also consider an alphabet of \emph{relational symbols} $\mathcal{R}$ which are used to name relational definitions.
The central syntactic category in the language is \emph{goal}. In our case, there are five types of goals: \emph{unification} of terms,
conjunction and disjunction of goals, fresh variable introduction, and invocation of some relational definition. Thus, unification is used
as a constraint, and multiple constraints can be combined using conjunction, disjunction, and recursion.
The final syntactic category is a \emph{specification} $\mathcal{S}$. It consists of a set
of relational definitions and a top-level goal. A top-level goal represents a search procedure which returns a stream of substitutions for
the free variables of the goal. The definition for a set of free variables for both terms and goals is conventional;
as ``\lstinline|fresh|''
is the sole binding construct the definition is rather trivial. The language we defined is first-order, as goals can not be passed as parameters,
returned or constructed at run time.

We now informally describe how relational search works. As we said, a goal represents a search procedure. This procedure takes a \emph{state} as input and returns a
stream of states; a state (among other information) contains a substitution that maps semantic variables into the terms over semantic variables. Then five types of
scenarios are possible (depending on the type of the goal):

\begin{itemize}
\item Unification ``\lstinline|$t_1$ === $t_2$|'' unifies terms $t_1$ and $t_2$ in the context of the substitution in the current state. If terms are unifiable,
  then their MGU is integrated into the substitution, and a one-element stream is returned; otherwise the result is an empty stream.
\item Conjunction ``\lstinline|$g_1$ /\ $g_2$|'' applies $g_1$ to the current state and then applies $g_2$ to each element of the result, concatenating
  the streams.
\item Disjunction ``\lstinline|$g_1$ \/ $g_2$|'' applies both its goals to the current state independently and then concatenates the results.
\item Fresh construct ``\lstinline|fresh $x$ . $g$|'' allocates a new semantic variable $\alpha$, substitutes all free occurrences of $x$ in $g$ with $\alpha$, and
  runs the goal.
\item Invocation ``$\lstinline|$R_i^{k_i}$ ($t_1$,...,$t_{k_i}$)|$'' finds a definition for the relational symbol \mbox{$R_i^{k_i}=\lambda x_1\dots x_{k_i}\,.\,g_i$}, substitutes
  all free occurrences of a formal parameter $x_j$ in $g_i$ with term $t_j$ (for all $j$) and runs the goal in the current state.
\end{itemize}

We stipulate that the top-level goal is preceded by an implicit ``\lstinline|fresh|'' construct, which binds all its free variables, and that the final substitutions
for these variables constitute the result of the goal evaluation.

Conjunction and disjunction form a monadic~\cite{Monads} interface with conjunction playing role of ``\lstinline|bind|'' and disjunction the role of ``\lstinline|mplus|''.
In this description, we swept a lot of important details under the carpet~--- for example, in actual implementations the components of disjunction are not evaluated in
isolation, but both disjuncts are evaluated incrementally with the control passing from one disjunct to another (\emph{interleaving})~\cite{Search};
the evaluation of some goals can be additionally deferred (via so-called ``\emph{inverse-$\eta$-delay}'')~\cite{MicroKanren}; instead of streams
the implementation can be based on ``ferns''~\cite{BottomAvoiding} to defer divergent computations, etc. In the following sections, we present
a complete formal description of relational semantics which resolves these uncertainties in a conventional way.

As an example consider the following specification. For the sake of brevity we
abbreviate immediately nested ``\lstinline|fresh|'' constructs into the one, writing ``\lstinline|fresh $x$ $y$ $\dots$ . $g$|'' instead of
``\lstinline|fresh $x$ . fresh $y$ . $\dots$ $g$|''.

\begin{tabular}{p{5.5cm}p{5.5cm}}
\begin{lstlisting}
append$^o$ = fun x y xy .
 ((x === Nil) /\ (xy === y)) \/
 (fresh h t ty .
   (x  === Cons (h, t))  /\
   (xy === Cons (h, ty)) /\
   (append$^o$ t y ty));

revers$^o$ x x
\end{lstlisting} &
\begin{lstlisting}
revers$^o$ = fun x xr .
 ((x === Nil) /\ (xr === Nil)) \/
 (fresh h t tr .
   (x === Cons (h, t)) /\
   (append$^o$ tr (Cons (h, Nil)) xr) /\
   (revers$^o$ t tr));
\end{lstlisting}
\end{tabular}

Here we defined\footnote{We respect here a conventional tradition for \textsc{miniKanren} programming to superscript all relational names with ``$^o$''.}
two relational symbols~--- ``\lstinline|append$^o$|'' and ``\lstinline|revers$^o$|'',~--- and specified a top-level goal ``\lstinline|revers$^o$ x x|''.
The symbol ``\lstinline|append$^o$|'' defines a relation of concatenation of lists~--- it takes three arguments and performs a case analysis on the first one. If the
first argument is an empty list (``\lstinline|Nil|''), then the second and the third arguments are unified. Otherwise, the first argument is deconstructed into a head ``\lstinline|h|''
and a tail ``\lstinline|t|'', and the tail is concatenated with the second argument using a recursive call to ``\lstinline|append$^o$|'' and additional variable ``\lstinline|ty|'', which
represents the concatenation of ``\lstinline|t|'' and ``\lstinline|y|''. Finally, we unify ``\lstinline|Cons (h, ty)|'' with ``\lstinline|xy|'' to form a final constraint. Similarly,
``\lstinline|revers$^o$|'' defines relational list reversing. The top-level goal represents a search procedure for all lists ``\lstinline|x|'', which are stable under reversing, i.e.
palindromes. Running it results in an infinite stream of substitutions:

\begin{lstlisting}
   $\alpha\;\mapsto\;$ Nil
   $\alpha\;\mapsto\;$ Cons ($\beta_0$, Nil)
   $\alpha\;\mapsto\;$ Cons ($\beta_0$, Cons ($\beta_0$, Nil))
   $\alpha\;\mapsto\;$ Cons ($\beta_0$, Cons ($\beta_1$, Cons ($\beta_0$, Nil)))
   $\dots$
\end{lstlisting}

where ``$\alpha$'' is a \emph{semantic} variable, corresponding to ``\lstinline|x|'', ``$\beta_i$'' are free semantic variables. Therefore, each substitution represents a set of all palindromes of a certain length.

\section{Denotational Semantics}
\label{denotational}

In this section, we present a denotational semantics for the language we defined above. We use a simple set-theoretic
approach analogous to the least Herbrand model for definite logic programs~\cite{LHM}.
Strictly speaking, instead of developing it from scratch we could have just described the conversion of specifications
into definite logic form and took their least Herbrand model. However, in that case, we would still need to define
the least Herbrand model semantics for definite logic programs in a certified way. In addition, while for
this concrete language the conversion to definite logic form is trivial, it may become less trivial for
its extensions (with, for example, nominal constructs~\cite{AlphaKanren}) which we plan to do in future.

We also must make the following observations. First, building inductive denotational semantics in a conventional way amounts to
constructing a complete lattice and a monotone function and taking its least fixed point~\cite{TarskiKnaster}.
As we deal with a first-order language with only monotonic constructs (conjunction/disjunction) these steps
are trivial. Moreover, we express the semantics in \textsc{Coq}, where all well-formed inductive definitions already
have proper semantics, which removes the necessity to justify the validity of the steps we perform. Second, 
the least Herbrand model is traditionally defined as the least fixed point of a transition function (defined by a logic program)
which maps sets of ground atoms to sets of ground atoms. We are, however, interested in \emph{relational} semantics which
should map a program into $n$-ary relation over ground terms, where $n$ is the number of free variables in the topmost
goal. Thus, we deviate from the traditional route and describe the denotational semantics in a more specific way.

To motivate further development, we first consider the following example. Let us have the following goal:

\begin{lstlisting}
   x === Cons (y, z)
\end{lstlisting}

There are three free variables, and solving the goal delivers us the following single answer:

\begin{lstlisting}
   $\alpha\mapsto\;$ Cons ($\beta$, $\gamma$)
\end{lstlisting}

where semantic variables $\alpha$, $\beta$ and $\gamma$ correspond to the syntactic ones ``\lstinline|x|'', ``\lstinline|y|'', ``\lstinline|z|''. The
goal does not put any constraints on ``\lstinline|y|'' and ``\lstinline|z|'', so there are no bindings for ``$\beta$'' and ``$\gamma$'' in the answer.
This answer can be seen as the following ternary relation over the set of all ground terms:

\[
\{(\mbox{\lstinline|Cons ($\beta$, $\,\gamma$)|}, \beta, \gamma) \mid \beta\in\mathcal{D},\,\gamma\in\mathcal{D}\}\subseteq\mathcal{D}^3
\]

The order of ``dimensions'' is important, since each dimension corresponds to a certain free variable. Our main idea is to represent this relation as a set of total
functions 

\[
\mathfrak{f}:\mathcal{A}\mapsto\mathcal{D}
\]

from semantic variables to ground terms. We call these functions \emph{representing functions}. Thus, we may reformulate the same relation as

\[
\{(\mathfrak{f}\,(\alpha),\mathfrak{f}\,(\beta),\mathfrak{f}\,(\gamma))\mid\mathfrak{f}\in\sembr{\mbox{\lstinline|$\alpha$ === Cons ($\beta$, $\,\gamma$)|}}\}
\]

where we use conventional semantic brackets ``$\sembr{\bullet}$'' to denote the semantics. For the top-level goal, we need to substitute its free syntactic
variables with distinct semantic ones, calculate the semantics, and build the explicit representation for the relation as shown above. The relation, obviously,
does not depend on the concrete choice of semantic variables but depends on the order in which the values of representing functions are tupled. This order can be
conventionalized, which gives us a completely deterministic semantics.

Now we implement this idea. First, for a representing function

\[
\mathfrak{f} : \mathcal{A}\to\mathcal{D}
\]

we introduce its homomorphic extension 

\[
  \overline{\mathfrak{f}}:\mathcal{T_A}\to\mathcal{D}
\]

which maps terms to terms:

\[
\begin{array}{rcl}
  \overline{\mathfrak f}\,(\alpha) & = & \mathfrak f\,(\alpha)\\
  \overline{\mathfrak f}\,(C_i^{k_i}\,(t_1,\dots.t_{k_i})) & = & C_i^{k_i}\,(\overline{\mathfrak f}\,(t_1),\dots \overline{\mathfrak f}\,(t_{k_i}))
\end{array}
\]

Let us have two terms $t_1, t_2\in\mathcal{T_A}$. If there is a unifier for $t_1$ and $t_2$ then, clearly, there is a substitution $\theta$ which
turns both $t_1$ and $t_2$ into the same \emph{ground} term (we do not require $\theta$ to be the most general). Thus, $\theta$ maps
(some) variables into ground terms, and its application to $t_{1(2)}$ is exactly $\overline{\theta}(t_{1(2)})$. This reasoning can be
performed in the opposite direction: a unification $t_1\equiv t_2$ defines the set of all representing functions $\mathfrak{f}$ for which
$\overline{\mathfrak{f}}(t_1)=\overline{\mathfrak{f}}(t_2)$.

We will use the conventional notions of pointwise modification of a function $f\,[x\gets v]$
and substitution $g\,[t/x]$ of a free variable $x$ with a term $t$ in a goal (or a term) $g$.


For a representing function $\mathfrak{f}:\mathcal{A}\to\mathcal{D}$ and a semantic variable $\alpha$ we define
the following \emph{generalization} operation:

\[
\mathfrak{f}\uparrow\alpha = \{ \mathfrak{f}\,[\alpha\gets d] \mid d\in\mathcal D\}
\]

Informally, this operation generalizes a representing function into a set of representing functions in such a way that the
values of these functions for a given variable cover the whole $\mathcal{D}$. We extend the generalization operation for sets of
representing functions $\mathfrak{F}\subseteq\mathcal{A}\to\mathcal{D}$:

\[
  \mathfrak{F}\uparrow\alpha = \bigcup_{\mathfrak{f}\in\mathfrak{F}}(\mathfrak{f}\uparrow\alpha)
\]

Now we are ready to specify the semantics for goals (see Fig.~\ref{denotational_semantics_of_goals}).
We've already given the motivation for
the semantics of unification: the condition $\overline{\mathfrak{f}}(t_1)=\overline{\mathfrak{f}}(t_2)$ gives us the set of all (otherwise
  unrestricted) representing functions which ``equate'' terms $t_1$ and $t_2$.
  Set union and intersection provide a conventional interpretation
for disjunction and conjunction of goals. In the case of a relational invocation we unfold the definition of the corresponding
relational symbol and substitute its formal parameters with actual ones.

The only non-trivial case is that of ``\lstinline|fresh $x$ . $g$|''. First, we take an arbitrary semantic variable $\alpha$,
not free in $g$, and substitute $x$ with $\alpha$. Then we calculate the semantics of $g\,[\alpha/x]$. The interesting part is the next step:
as $x$ can not be free in ``\lstinline|fresh $x$ . $g$|'', we need to generalize the result over $\alpha$ since in our model the semantics of a
goal specifies a relation over its free variables. We introduce some nondeterminism by choosing arbitrary $\alpha$, but we can prove that with different
choices of free variable the semantics of a goal does not change.

\begin{lemma}
\label{lem:den_sem_change_var}
For any goal \lstinline|fresh $x$ . $g$|, for any two variables $\alpha$ and $\beta$ which are not free in this goal,
if $\mathfrak{f} \in \sembr{g\,[\alpha/x]}$, then for any representing function $\mathfrak{f}'$, such that

\begin{enumerate}
\item $\mathfrak{f}'(\beta) = \mathfrak{f}(\alpha)$
\item $\forall \gamma: \gamma \neq \alpha \land \gamma \neq \beta,\; \mathfrak{f}'(\gamma) = \mathfrak{f}(\gamma)$
\end{enumerate}

\noindent it is true that $\mathfrak{f}' \in \sembr{g\,[\beta/x]}$.
\end{lemma}
  The proof turned out to be the most cumbersome among all others in the case where $g$ is a nested \lstinline|fresh| contruct. In that case, we have to constructively build two representing functions (including an intermediate one for an intermediate goal) by pointwise modification. The details of this proof can be found in the Appendix~\ref{appendix_den_sem_change_var_proof}, the full proof script is in the specification in Coq.

\begin{figure}[t]
  \[
  \begin{array}{cclr}
    \sembr{t_1\equiv t_2}&=&\{\mathfrak f : \mathcal{A}\to\mathcal{D}\mid \overline{\mathfrak{f}}\,(t_1)=\overline{\mathfrak{f}}\,(t_2)\}& \ruleno{Unify$_D$}\\
    \sembr{g_1\wedge g_2}&=&\sembr{g_1}\cap\sembr{g_2}&\ruleno{Conj$_D$}\\
    \sembr{g_1\vee g_2}&=&\sembr{g_1}\cup\sembr{g_2}&\ruleno{Disj$_D$}\\
    \sembr{\mbox{\lstinline|fresh|}\,x\,.\,g}&=&(\sembr{g\,[\alpha/x]})\uparrow\alpha,\;\alpha\not\in FV(g)& \ruleno{Fresh$_D$}\\
    \sembr{R\,(t_1,\dots,t_k)}&=&\sembr{g\,[t_1/x_1,\dots,t_k/x_k]},\;\mbox{where}\;R=\lambda\,x_1\dots x_k\,.\,g & \ruleno{Invoke$_D$}
  \end{array}
  \]
  \caption{Denotational semantics of goals}
  \label{denotational_semantics_of_goals}
\end{figure}

We can prove the following important \emph{closedness condition} for the semantics of a goal $g$.

\begin{lemma}[Closedness condition]
\label{lem:closedness_condition}
For any goal $g$ and two representing functions ${\mathfrak f}$ and ${\mathfrak f'}$, such that $\left.{\mathfrak f}\right|_{FV(g)} = \left.{\mathfrak f'}\right|_{FV(g)}$, it is true, that
${\mathfrak f} \in \sembr{g} \Leftrightarrow {\mathfrak f'} \in \sembr{g}$.
\end{lemma}

In other words, representing functions for a goal $g$ restrict only the values of free variables of $g$ and do not introduce any ``hidden'' correlations.
This condition guarantees that our semantics is closed in the sense that it does not introduce artificial restrictions for the relation it defines.

\section{Operational Semantics}
\label{operational}

In this section we describe the operational semantics of \textsc{miniKanren}, which corresponds to the known
implementations with interleaving search. The semantics is given in the form of a labeled transition system (LTS)~\cite{LTS}. From now on we
assume the set of semantic variables to be linearly ordered ($\mathcal{A}=\{\alpha_1,\alpha_2,\dots\}$).

We introduce the notion of substitution

\[
  \sigma : \mathcal{A}\to\mathcal{T_A}
\]

\noindent as a (partial) mapping from semantic variables to terms over the set of semantic variables. We denote $\Sigma$ the
set of all substitutions, $\dom{\sigma}$~--- the domain for a substitution $\sigma$,
$\ran{\sigma}=\bigcup_{\alpha\in\mathcal{D}om\,(\sigma)}\fv{\sigma\,(\alpha)}$~--- its range (the set of all free variables in the image).

The \emph{non-terminal states} in the transition system have the following shape:

\[
S = \mathcal{G}\times\Sigma\times\mathbb{N}\mid S\oplus S \mid S \otimes \mathcal{G}
\]

As we will see later, an evaluation of a goal is separated into elementary steps, and these steps are performed interchangeably for different subgoals. 
Thus, a state has a tree-like structure with intermediate nodes corresponding to partially-evaluated conjunctions (``$\otimes$'') or
disjunctions (``$\oplus$''). A leaf in the form $\inbr{g, \sigma, n}$ determines a goal in a context, where $g$ is a goal, $\sigma$ is a substitution accumulated so far,
and $n$ is a natural number, which corresponds to a number of semantic variables used to this point. For a conjunction node, its right child is always a goal since
it cannot be evaluated unless some result is provided by the left conjunct.

The full set of states also include one separate terminal state (denoted by $\diamond$), which symbolizes the end of the evaluation.

\[
\hat{S} = \diamond \mid S
\]

We will operate with the well-formed states only, which are defined as follows.

\begin{definition}
  Well-formedness condition for extended states:
  
  \begin{itemize}
  \item $\diamond$ is well-formed;
  \item $\inbr{g, \sigma, n}$ is well-formed iff $\fv{g}\cup\dom{\sigma}\cup\ran{\sigma}\subseteq\{\alpha_1,\dots,\alpha_n\}$;
  \item $s_1\oplus s_2$ is well-formed iff $s_1$ and $s_2$ are well-formed;
  \item $s\otimes g$ is well-formed iff $s$ is well-formed and for all leaf triplets $\inbr{\_,\_,n}$ in $s$ it is true that $\fv{g}\subseteq\{\alpha_1,\dots,\alpha_n\}$.
  \end{itemize}
  
\end{definition}

Informally the well-formedness restricts the set of states to those in which all goals use only allocated variables.

Finally, we define the set of labels:

\[
L = \step \mid \Sigma\times \mathbb{N}
\]

The label ``$\step$'' is used to mark those steps which do not provide an answer; otherwise, a transition is labeled by a pair of a substitution and a number of allocated
variables. The substitution is one of the answers, and the number is threaded through the derivation to keep track of allocated variables.

\begin{figure*}[t]
  \renewcommand{\arraystretch}{1.6}
  \[
  \begin{array}{cr}
    \inbr{t_1 \equiv t_2, \sigma, n} \xrightarrow{\step} \Diamond , \, \, \nexists\; mgu\,(t_1 \sigma, t_2 \sigma) &\ruleno{UnifyFail} \\
    \inbr{t_1 \equiv t_2, \sigma, n} \xrightarrow{(mgu\,(t_1 \sigma, t_2 \sigma) \circ \sigma,\, n)} \Diamond & \ruleno{UnifySuccess} \\
    \inbr{g_1 \lor g_2, \sigma, n} \xrightarrow{\step} \inbr{g_1, \sigma, n} \oplus \inbr{g_2, \sigma, n} & \ruleno{Disj} \\
    \inbr{g_1 \land g_2, \sigma, n} \xrightarrow{\step} \inbr{ g_1, \sigma, n} \otimes g_2 & \ruleno{Conj} \\
    \inbr{\mbox{\lstinline|fresh|}\, x\, .\, g, \sigma, n} \xrightarrow{\step} \inbr{g\,[\bigslant{\alpha_{n + 1}}{x}], \sigma, n + 1} & \ruleno{Fresh} \\
    \dfrac{R_i^{k_i}=\lambda\,x_1\dots x_{k_i}\,.\,g}{\inbr{R_i^{k_i}\,(t_1,\dots,t_{k_i}),\sigma,n} \xrightarrow{\step} \inbr{g\,[\bigslant{t_1}{x_1}\dots\bigslant{t_{k_i}}{x_{k_i}}], \sigma, n}} & \ruleno{Invoke}\\
    \dfrac{s_1 \xrightarrow{\step} \Diamond}{(s_1 \oplus s_2) \xrightarrow{\step} s_2} & \ruleno{SumStop}\\
    \dfrac{s_1 \xrightarrow{r} \Diamond}{(s_1 \oplus s_2) \xrightarrow{r} s_2} & \ruleno{SumStopAns}\\
    \dfrac{s \xrightarrow{\step} \Diamond}{(s \otimes g) \xrightarrow{\step} \Diamond} &\ruleno{ProdStop}\\
    \dfrac{s \xrightarrow{(\sigma, n)} \Diamond}{(s \otimes g) \xrightarrow{\step} \inbr{g, \sigma, n}}  & \ruleno{ProdStopAns}\\
    \dfrac{s_1 \xrightarrow{\step} s'_1}{(s_1 \oplus s_2) \xrightarrow{\step} (s_2 \oplus s'_1)} &\ruleno{SumStep}\\
    \dfrac{s_1 \xrightarrow{r} s'_1}{(s_1 \oplus s_2) \xrightarrow{r} (s_2 \oplus s'_1)} &\ruleno{SumStepAns}\\
    \dfrac{s \xrightarrow{\step} s'}{(s \otimes g) \xrightarrow{\step} (s' \otimes g)} &\ruleno{ProdStep}\\
    \dfrac{s \xrightarrow{(\sigma, n)} s'}{(s \otimes g) \xrightarrow{\step} (\inbr{g, \sigma, n} \oplus (s' \otimes g))} & \ruleno{ProdStepAns} 
  \end{array}
  \]
  \caption{Operational semantics of interleaving search}
  \label{lts}
\end{figure*}

The transition rules are shown in Fig.~\ref{lts}. The first two rules specify the semantics of unification. If two terms are not unifiable under the current substitution
$\sigma$ then the evaluation stops with no answer; otherwise, it stops with the most general unifier applied to a current substitution as an answer.

The next two rules describe the steps performed when disjunction or conjunction is encountered on the top level of the current goal. For disjunction, it schedules both goals (using ``$\oplus$'') for
evaluating in the same context as the parent state, for conjunction~--- schedules the left goal and postpones the right one (using ``$\otimes$'').

The rule for ``\lstinline|fresh|'' substitutes bound syntactic variable with a newly allocated semantic one and proceeds with the goal.

The rule for relation invocation finds a corresponding definition, substitutes its formal parameters with the actual ones, and proceeds with the body.

The rest of the rules specify the steps performed during the evaluation of two remaining types of the states~--- conjunction and disjunction. In all cases, the left state
is evaluated first. If its evaluation stops, the disjunction evaluation proceeds with the right state, propagating the label (\textsc{SumStop} and \textsc{SumStep}), and the conjunction schedules the right goal for evaluation in the context of the returned answer (\textsc{ProdStopAns}) or stops if there is no answer (\textsc{ProdStop}).

The last four rules describe \emph{interleaving}, which occurs when the evaluation of the left state suspends with some residual state (with or without an answer). In the case of disjunction
the answer (if any) is propagated, and the constituents of the disjunction are swapped (\textsc{SumStep}, \textsc{SumStepAns}). In the case of conjunction, if the evaluation step in
the left conjunct did not provide any answer, the evaluation is continued in the same order since there is still no information to proceed with the evaluation of the right
conjunct (\textsc{ProdStep}); if there is some answer, then the disjunction of the right conjunct in the context of the answer and the remaining conjunction is
scheduled for evaluation (\textsc{ProdStepAns}).

The introduced transition system is completely deterministic: there is exactly one transition from any non-terminal state.
There is, however, some freedom in choosing the order of evaluation for conjunction and
disjunction states. For example, instead of evaluating the left substate first, we could choose to evaluate the right one, etc.
This choice reflects the inherent non-deterministic nature of search in relational (and, more generally, logical) programming.
Although we could introduce this ambiguity into the semantics (by replacing specific rules for disjunctions and conjunctions evaluation with some conditions on it), we want an operational semantics that would be easy to present and easy to employ to describe existing language extensions (already described for a specific implementation of interleaving search), so we instead base the semantics on one canonical search strategy.
At the same time, as long as deterministic search procedures are sound and complete, we can consider them ``equivalent''.\footnote{There still can be differences in observable behavior of concrete goals under different sound and complete search strategies.
For example, a goal can be refutationally complete~\cite{WillThesis} under one strategy and non-complete under another.}

It is easy to prove that transitions preserve well-formedness of states.

\begin{lemma}{(Well-formedness preservation)}
\label{lem:well_formedness_preservation}
For any transition $s \xrightarrow{l} \hat{s}$, if $s$ is well-formed then $\hat{s}$ is also well-formed.
\end{lemma}

A derivation sequence for a certain state determines a \emph{trace}~--- a finite or infinite sequence of answers. The trace corresponds to the stream of answers
in the reference \textsc{miniKanren} implementations. We denote a set of answers in the trace for state $\hat{s}$ by $\tr{\hat{s}}$.

We can relate sets of answers for the partially evaluated conjunction and disjunction with sets of answers for their constituents by the two following lemmas.

\begin{lemma}
\label{lem:sum_answers}
For any non-terminal states $s_1$ and $s_2$, $\tr{s_1 \oplus s_2} = \tr{s_1} \cup \tr{s_2}$.
\end{lemma}

\begin{lemma}
\label{lem:prod_answers}
For any non-terminal state $s$ and goal $g$,  \mbox{$\tr{s \otimes g} \supseteq \bigcup_{(\sigma, n) \in \tr{s}} \tr{\inbr{g, \sigma, n}}$}.
\end{lemma}

These two lemmas constitute the exact conditions on definition of these operators that we will use to prove the completeness of an operational semantics.

We also can easily describe the criterion of termination for disjunctions.

\begin{lemma}
\label{lem:disj_termination}
For any goals $g_1$ and $g_2$, sunbstitution $\sigma$, and number $n$, the trace from the state $\inbr{g_1 \vee g_2, \sigma, n}$ is finite iff the traces from both $\inbr{g_1, \sigma, n}$ and $\inbr{g_2, \sigma, n}$ are finite.
\end{lemma}

These simple statements already allow us to prove two important properties of interleaving search as corollaries: the ``fairness'' of disjunction~--- the fact that the trace for disjunction contains all the answers from both streams for disjuncts~--- and the ``commutativity'' of disjunctions~--- the fact that swapping two disjuncts (at the top level) does not change the termination of the goal evaluation. 

\section{Equivalence of Semantics}
\label{equivalence}

Now we can relate two different kinds of semantics for \textsc{miniKanren} described in the previous sections and show that the results given by these two semantics are the same for any specification.
This will actually say something important about the search in the language: since operational semantics describes precisely the behavior of the search and denotational semantics
ignores the search and describes what we \emph{should} get from a mathematical point of view, by proving their equivalence we establish the \emph{completeness} of the search, which
means that the search will get all answers satisfying the described specification and only those.

But first, we need to relate the answers produced by these two semantics as they have different forms: a trace of substitutions (along with the numbers of allocated variables)
for the operational one and a set of representing functions for the denotational one. We can notice that the notion of a representing function is close to substitution, with only two differences:

\begin{itemize}
\item representing functions are total;
\item terms in the domain of representing functions are ground.
\end{itemize}

Therefore we can easily extend (perhaps ambiguously) any substitution to a representing function by composing it with an arbitrary representing function preserving
all variable dependencies in the substitution. So we can define a set of representing functions that correspond to a substitution as follows:

\[
\sembr{\sigma} = \{\overline{\mathfrak f} \circ \sigma \mid \mathfrak{f}:\mathcal{A}\mapsto\mathcal{D}\}
\]

And the \emph{denotational analog} of operational semantics (a set of representing functions corresponding to the answers in the trace) for a given state $\hat{s}$ is
then defined as the union of sets for all substitutions in the trace:

\[
\sembr{\hat{s}}_{op} = \cup_{(\sigma, n) \in \tr{\hat{s}}} \sembr{\sigma}
\]

This allows us to state theorems relating the two semantics.

\begin{theorem}[Operational semantics soundness]
\label{lem:soundness}
If indices of all free variables in a goal $g$ are limited by some number $n$, then $\sembr{\inbr{g, \epsilon, n}}_{op} \subseteq \sembr{g}$.
\end{theorem}

It can be proven by nested induction, but first, we need to generalize the statement so that the inductive hypothesis is strong enough for the inductive step.
To do so, we define denotational semantics not only for goals but for arbitrary states. Note that this definition does not need to have any intuitive
interpretation, it is introduced only for the proof to go smoothly. The definition of the denotational semantics for extended states is shown on Fig.~\ref{denotational_semantics_of_states}.
The generalized version of the theorem uses it.

\begin{figure}[t]
  \[
  \begin{array}{ccl}
    \sembr{\Diamond}&=&\emptyset\\
    \sembr{\inbr{g, \sigma, n}}&=&\sembr{g}\cap\sembr{\sigma}\\
    \sembr{s_1 \oplus s_2}&=&\sembr{s_1}\cup\sembr{s_2}\\
    \sembr{s \otimes g}&=&\sembr{s}\cap\sembr{g}\\
  \end{array}
  \]
  \caption{Denotational semantics of states}
  \label{denotational_semantics_of_states}
\end{figure}

\begin{lemma}[Generalized soundness]
\label{lem:gen_soundness}
For any well-formed state $\hat{s}$

\[
\sembr{\hat{s}}_{op} \subseteq \sembr{\hat{s}}.
\]
\end{lemma}

It can be proven by the induction on the number of steps in which a given answer (more accurately, the substitution that contains it) occurs in the trace.
We break the proof in two parts and separately prove by induction on evidence that for every transition in our system the semantics of both the label (if there is one)
and the next state are subsets of the denotational semantics for the initial state.

\begin{lemma}[Soundness of the answer]
\label{lem:answer_soundness}
For any transition $s \xrightarrow{(\sigma, n)} \hat{s}$, \mbox{$\sembr{\sigma} \subseteq \sembr{s}$}.
\end{lemma}

\begin{lemma}[Soundness of the next state]
\label{lem:next_state_soundness}
For any transition $s \xrightarrow{l} \hat{s}$, \mbox{$\sembr{\hat{s}} \subseteq \sembr{s}$}.
\end{lemma}

It would be tempting to formulate the completeness of operational semantics as soundness with the inverted inclusion, but it does not hold in such generality.
The reason for this is that the denotational semantics encodes only the dependencies between free variables of a goal, which is reflected by the closedness condition,
while the operational semantics may also contain dependencies between semantic variables allocated in \lstinline|fresh| constructs. Therefore we formulate completeness
with representing functions restricted on the semantic variables allocated in the beginning (which includes all free variables of a goal). This does not
compromise our promise to prove the completeness of the search as \textsc{miniKanren} returns substitutions only for queried variables,
which are allocated in the beginning.

\begin{theorem}[Operational semantics completeness]
If the indices of all free variables in a goal $g$ are limited by some number $n$, then

\[
\{\mathfrak{f}|_{\{\alpha_1,\dots,\alpha_n\}} \mid \mathfrak{f} \in \sembr{g}\} \subseteq \{\mathfrak{f}|_{\{\alpha_1,\dots,\alpha_n\}} \mid \mathfrak{f} \in \sembr{\inbr{g, \epsilon, n}}_{op}\}.
\]
\end{theorem}

Similarly to the soundness, this can be proven by nested induction, but the generalization is required. This time it is enough to generalize it from goals
to states of the shape $\inbr{g, \sigma, n}$. We also need to introduce one more auxiliary semantics~--- \emph{step-indexed denotational semantics} (denoted by $\sembr{\bullet}^i$). It is an implementation of the well-known approach~\cite{StepIndexing} of indexing typing or semantic logical relations by a number of permitted evaluation steps to allow inductive reasoning on it.
In our case, $\sembr{g}^i$ includes only those representing functions that one can get after no more than $i$ unfoldings of relational calls.

The step-indexed denotational semantics is an approximation of the conventional denotational semantics; it is clear that any answer in conventional denotational semantics will also be in step-indexed denotational semantics for some number of steps.

\begin{lemma}
$\sembr{g} \subseteq \cup_i \sembr{g}^i$
\end{lemma}

Now the generalized version of the completeness theorem is as follows.

\begin{lemma}[Generalized completeness]
\label{lem:gen_completeness}
For any set of relational definitions, for any number of unfoldings $i$, for any well-formed state $\inbr{g, \sigma, n}$,

\[
\{\mathfrak{f}|_{\{\alpha_1,\dots,\alpha_n\}} \mid \mathfrak{f} \in \sembr{g}^i \cap \sembr{\sigma}\} \subseteq \{\mathfrak{f}|_{\{\alpha_1,\dots,\alpha_n\}} \mid \mathfrak{f} \in \sembr{\inbr{g, \sigma, n}}_{op}\}.
\]
\end{lemma}

The proof is by the induction on nuber of unfoldings $i$. The induction step is proven by structural induction on goal $g$. We use lemmas~\ref{lem:sum_answers} and~\ref{lem:prod_answers} for evaluation of a disjunction and a conjunction respectively, and lemma~\ref{lem:den_sem_change_var} in the case of fresh variable introduction to move from an arbitrary semantic variable in denotational semantics to the next allocated fresh variable. The details of this proof may be found in Appendix~\ref{appendix_gen_completeness_proof}, the full proof script is in the specification in Coq.

\section{Specification in \textsc{Coq}}
\label{specification}

We certified all the definitions and propositions from the previous sections using the \textsc{Coq} proof assistant.\footnote{The specification is available at \url{https://github.com/dboulytchev/miniKanren-coq}} The \textsc{Coq} specification for the most part closely follows the formal descriptions we gave by means of inductive definitions (and inductively defined propositions in particular) and structural induction in proofs. The detailed description of the specification, including code snippets, is provided in Appendix~\ref{appendix_coq}, and in this section we address only some non-trivial parts of it and some design choices.

The language formalized in \textsc{Coq} has a few non-essential simplifications for the sake of convenience. Specifically, we restrict the arities of all constructors to be either zero or two and require all relations to have exactly one argument. These restrictions do not make the language less expressive in any way since we can always represent a sequence of terms as a list using constructors \lstinline|Nil$^0$| and \lstinline|Cons$^2$|. 

In our formalization of the language we use higher-order abstract syntax~\cite{HOAS} for variable binding, therefore we work explicitly only with semantic variables. We preferred it to the first-order syntax because it gives us the ability to use substitution and the induction principle provided by \textsc{Coq}. On the other hand, we need to explicitly specify a requirement on the syntax representation, which is trivially fulfilled in the first-order case: all bindings have to be ``consistent'', i.e. if we instantiate a higher-order \lstinline|fresh| construct with different semantic variables the results will be the same up to some renaming (provided that both those variables are not free in the body of the binder). Another requirement we have to specify explicitly (independent of HOAS/FOAS dichotomy) is a requirement that the definitions of relations do not contain unbound semantic variables.

To formalize the operational semantics in \textsc{Coq} we first need to define all preliminary notions from unification theory~\cite{Unification} which our semantics uses. In particular, we need to implement the notion of the most general unifier (MGU). As it is well-known~\cite{StructuralMGU} all standard recursive algorithms for calculating MGU are not decreasing on argument terms, so we can't define them as simple recursive functions in \textsc{Coq} due to the termination check failure. The standard approach to tackle this problem is to define the function through well-founded recursion. We use a distinctive version of this approach, which is more convenient for our purposes: we define MGU as a proposition (for which there is no termination requirement in \textsc{Coq}) with a dedicated structurally-recursive function for one step of unification, and then we use a well-founded induction to prove the existence of a corresponding result for any arguments and defining properties of MGU. For this well-founded induction, we use the number of distinct free variables in argument terms as a well-founded order on pairs of terms.

In the operational semantics, to define traces as (possibly) infinite sequences of transitions we use the standard approach in \textsc{Coq}~--- coinductively defined streams. Operating with them requires a number of well-known tricks, described by Chlipala~\cite{CPDT}, to be applied, such as the use of a separate coinductive definition of equality on streams.

The final proofs of soundness and completeness of operational semantics are relatively small, but the large amount of work is hidden in the proofs of auxiliary facts that they use (including lemmas from the previous sections and some technical machinery for handling representing functions).

\section{Applications}
\label{applications}

In this section, we consider some applications of the framework and results, described in the previous sections.

\subsection{Correctness of Transformations}

One important immediate corollary of the equivalence theorems we have proven is the justification of correctness for certain program transformations.
The completeness of interleaving search guarantees the correctness of any transformation that preserves the denotational semantics,
for example:

\begin{itemize}
\item changing the order of constituents in conjunctions and disjunctions;
\item distributing conjunctions over disjunctions and vice versa, for example, normalizing goals info CNF or DNF;
\item moving fresh variable introduction upwards/downwards, for example, transforming any relation into a top-level fresh
  construct with a freshless body.
\end{itemize}

Note that this way we can guarantee only the preservation of results as \emph{sets of ground terms}; the other aspects of program behavior,
such as termination, may be affected by some of these transformations.\footnote{Possible slowdown and loss of termination after reorderings in conjunction is a famous example of this phenomenon in \textsc{miniKanren}, known as conjunction non-commutativity~\cite{WillThesis}.}

One of the applications for these transformations is a conversion from/to \textsc{Prolog}. As both languages use essentially the same fragment of first-order logic,
their programs are mutually convertible. The conversion from \textsc{Prolog} to \textsc{miniKanren} is simpler as the latter admits a richer syntax of goals. The inverse
conversion involves the transformation into a DNF and splitting the disjunction into a number of separate clauses. This transformation, in particular, makes it possible to
reuse our approach to describe the semantics of \textsc{Prolog} as well. In the following sections we briefly address this problem.

\subsection{SLD Semantics}
\label{sld}

The conventional \textsc{Prolog} SLD search differs from the interleaving one in just one aspect~--- it does not perform interleaving.
Thus, changing just two rules in the operational semantics converts interleaving search into the depth-first one:

\[
  \begin{array}{crcr}
    \dfrac{s_1 \xrightarrow{\circ} s'_1}{(s_1 \oplus s_2) \xrightarrow{\circ} (s'_1 \oplus s_2)} &\ruleno{DisjStep}&
    \dfrac{s_1 \xrightarrow{r} s'_1}{(s_1 \oplus s_2) \xrightarrow{r} (s'_1 \oplus s_2)} &\ruleno{DisjStepAns}
  \end{array}
\]
\vskip3mm

With this definition we can almost completely reuse the mechanized proof of soundness (with minor changes); the completeness, however,
can no longer be proven (as it does not hold anymore).

\subsection{Cut}
\label{cut}

Dealing with the ``cut'' construct is known to be a cornerstone feature in the study of operational semantics for \textsc{Prolog}. It turned out that
in our case the semantics of ``cut'' can be expressed naturally (but a bit verbosely). Unlike SLD-resolution, it does not amount to an incremental
change in semantics description. It also would work only for programs directly converted from \textsc{Prolog} specifications.

The key observation in dealing with the ``cut'' in our setting is that a state in our semantics, in fact, encodes the whole current
search tree (including all backtracking possibilities). This opens the opportunity to organize proper ``navigation'' through the tree
to reflect the effect of ``cut''. The details of the semantic description can be found in the Appendix~\ref{appendix_cut}.

For this semantics, we can repeat the proof of soundness w.r.t. to the denotational semantics. There is, however, a little subtlety with our construction:
we cannot formally prove that our semantics indeed encodes the conventional meaning of ``cut'' (since we do not have other semantics of ``cut'' to compare with).
Nevertheless, we can demonstrate a plausible behavior using the extracted reference interpreter.

\subsection{Reference Interpreters}

Using the \textsc{Coq} extraction mechanism, we extracted two reference interpreters from our definitions and theorems: one for conventional
\textsc{miniKanren} with interleaving search and another one for SLD search with cut. These interpreters can be used to practically investigate the behavior
of specifications in unclear, complex, or corner cases. Our experience has shown that these interpreters demonstrate the expected behavior
in all cases.

\section{Related Work}

The study of formal semantics for logic programming languages, particularly \textsc{Prolog}, is a well-established research domain. Early
works~\cite{JonesMycroftSemantics,DebrayMishraSemantics} addressed the computational aspects of both pure \textsc{Prolog} and its extension
with the cut construct. Recently, the application of certified/mechanized approaches came into focus as well. In particular,
in one work~\cite{CertifiedPrologEquivalences} the equivalence of a few differently defined semantics
for pure \textsc{Prolog} is proven, and in another work~\cite{CeritfiedDenotationalCut} a denotational semantics for \textsc{Prolog} with cut is presented; both
works provide \textsc{Coq}-mechanized proofs. It is interesting that the former one also advocates the use of higher-order
abstract syntax. We are not aware of any prior work on certified semantics for \textsc{Prolog} which contributed a correct-by-construction
interpreter. Our certified description of SLD resolution with cut can be considered as a certified semantics for \textsc{Prolog} modulo
occurs check in unification (which \textsc{Prolog} does not have by default).

The implementation of first-order unification in dependently typed languages constitutes a well-known challenge with a number of
known solutions. The major difficulty comes from the non-structural recursivity of conventional unification algorithms, which
requires to provide a witness for convergence. The standard approach is to define a generally-recursive function and a well-founded order
for its arguments. This route is taken in a number of works~\cite{MGUinLCF,MGUinMLTT,IdempMGUinCoq,TextbookMGUinCoq}, where the descriptions of
unification algorithms are given in \textsc{Coq}, \textsc{LCF} and \textsc{Alf}. The well-founded used there is
lexicographically ordered tuples, containing the information about the number of different free variables and the sizes of
the arguments. We implement a similar approach, but we separate the test for the non-matching case into a dedicated
function. Thus, we make a recursive call only when the current substitution extension is guaranteed, which allows us to use the
number of different free variables as the well-founded order. An alternative approach suggested by McBride~\cite{StructuralMGU} gives a structurally recursive definition of
the unification algorithm; this is achieved by indexing the arguments with the numbers of their free variables.

The use of higher-order abstract syntax (HOAS) for dealing with language constructs in \textsc{Coq} was addressed in early work~\cite{HOASinCoq},
where it was employed to describe the lambda calculus. The inconsistency phenomenon of HOAS representation, mentioned in Section~\ref{specification}, is called
there ``exotic terms'' there and is handled using a dedicated inductive predicate ``\lstinline|Valid_v|''. The predicate has a non-trivial implementation based
on subtle observations on the behavior of bindings. Our case, however, is much simpler: there is not much variety in ``exotic terms'' (for example, we do not have
reductions in terms), and our consistency predicate can be considered as a limited version of ``\lstinline|Valid_v|'' for a bigger language.

The study of formal semantics for \textsc{miniKanren} is not a completely novel venture. Previously, a nondeterministic
small-step semantics was described~\cite{RelConversion}, as well as a big-step semantics for a finite number of answers~\cite{DivTest};
neither uses proof mechanization and in both works the interleaving is not addressed. 

The work of Kumar~\cite{MechanisingMiniKanren} can be considered as our direct predecessor. It also introduces both denotational and
operational semantics and presents a \textsc{HOL}-certified proof for the soundness of the latter w.r.t. the former. The denotational
semantics resembles ours but considers only queries with a single free variable (we do not see this restriction as important).
On the other hand, the operational semantics is non-deterministic, which makes it
impossible to express interleaving and extract the interpreter in a direct way. In addition, a specific form of ``executable semantics''
is introduced, but its connection to the other two is not established. Finally, no completeness result is presented.
We consider our completeness proof as an essential improvement. 

The most important property of interleaving search~--- completeness~--- was postulated in the introductory paper~\cite{Search}, and is delivered by
all major implementations. Hemann et al.~\cite{SmallEmbedding} give a proof of completeness for a specific implementation of \textsc{miniKanren};
however, the completeness is understood there as
preservation of all answers during the interleaving of answer streams, i.e. in a more narrow sense than in our work since no relation
to denotational semantics is established.

\section{Conclusion and Future Work}

In this paper, we presented a certified formal semantics for core \textsc{miniKanren} and proved some of its basic properties
(including interleaving search completeness, disjunction fairness and commutativity), which are believed to hold in existing implementations.
We also derived a semantics for conventional SLD resolution with cut and extracted two certified reference interpreters.
We consider our work as the initial setup for the future development of \textsc{miniKanren} semantics.

The language we considered here lacks many important features, which are already introduced
and employed in many implementations. Integrating these extensions~--- in the first hand, disequality constraints,~--- into
the semantics looks a natural direction for future work. We are also going to address the problems of proving some
properties of relational programs (equivalence, refutational completeness, etc.).



\bibliographystyle{abbrv} 
\bibliography{main}

\appendix
\clearpage

\section{Details of Proofs}
\label{appendix_proofs}

\subsection{Proof of Lemma~\ref{lem:den_sem_change_var}}
\label{appendix_den_sem_change_var_proof}

The proof goes by structural induction, all cases naturally use inductive hypotheses, except for the case of fresh variable introduction. We examine this case in more detail.

We have a nested fresh variable introduction $\mbox{\lstinline|fresh|}\,y\,.\,g'$ as the goal $g$. $g'$ may contain syntactic variables $x$ and $y$ (the case when $g$ does not contain syntactic variable $x$ is trivial and we have to consider it separately). By the statement of the lemma and the definition of denotational semantics we have ${\mathfrak f'_1} \in \sembr{g'\,[\alpha_1/x]\,[\alpha_3/y]}$, where $\alpha_3$ is some fresh semantic variable, and ${\mathfrak f'_1}$ is some representing function which may differ from ${\mathfrak f_1}$ only on variable $\alpha_3$. Variable $\alpha_3$ may not be equal to $\alpha_1$, but may be equal to $\alpha_2$ and further in proof we sometimes have to consider these two cases separetly, because they are essentially different. We should find some fresh variable $\alpha_4$ and representing function ${\mathfrak f'_2} \in \sembr{g'\,[\alpha_2/x]\,[\alpha_4/y]}$ that may differ from ${\mathfrak f_2}$ only on variable $\alpha_4$.

Among different options we considered, the easiest choice for a variable $\alpha_4$ turned out to be the freshest possible variable (one that is not equal to any of the mentioned above and that does not occur in the goal $g'$). Then we need to move from the function ${\mathfrak f'_1}$ to the function ${\mathfrak f'_2}$ by applying the induction hypothesis twice (since now we change two substituted variables simultaneously). To do this we need to carefully change the values of the function on a few suitable points. It turned out, that we have to change the values on all four mentioned variables. First, we build an intermediate function ${\mathfrak f'_{1.5}} = {\mathfrak f'_1}[\alpha_3\gets {\mathfrak f_2}(\alpha_3)][\alpha_4\gets {\mathfrak f'_1}(\alpha_3)]$ that should belong to an intermediate semantics $\sembr{g'\,[\alpha_1/x]\,[\alpha_4/y]}$. Then we build a final function ${\mathfrak f'_2} = {\mathfrak f'_{1.5}}[\alpha_1\gets {\mathfrak f_2}(\alpha_1)][\alpha_2\gets {\mathfrak f'_{1.5}}(\alpha_1)]$ on top of it. For both transitions, we carefully check that all conditions on two functions from the statement of the lemma (and, therefore, from the induction hypothesis) are satisfied by examination of values of the functions on all mentioned variables.

\subsection{Proof of Generalized Completeness}
\label{appendix_gen_completeness_proof}

The proof goes by induction on the number of unfoldings in step-indexed denotational semantics, then by structural induction on the goal.

\begin{itemize}
\item In case of unification $t_1\equiv t_2$ we need to show that any representing function that unifies terms $t_1$ and $t_2$ and is an extension of $\sigma$ also is an extension of $mgu\,(t_1 \sigma, t_2 \sigma) \circ \sigma$. This requires some observations about representing functions, most importantly the fact that for any representing function $f$, unifying two terms, there exists a unifying substitution for these terms, for which $f$ is an extension.
\item In case of disjunction we apply lemma~\ref{lem:sum_answers} to induction hypotheses.
\item In case of conjunction we apply lemma~\ref{lem:prod_answers} to induction hypotheses. Note that we use two nested inductions instead of one induction on evidence because of this case: we need to apply an induction hypothesis to different representing functions (which may have different values on non-allocated free variables), and otherwise the induction hypotheses would not be flexible enough.
\item In the case of fresh variable introduction, we can not simply use induction hypothesis, because it has a goal with an arbitrary fresh variable substituted (see definition of denotational semantics on Fig.~\ref{denotational_semantics_of_goals}), but we need to relate two semantics of the goal with exactly the first non-allocated variable substituted (from transition in operational semantics, see Fig.~\ref{lts}). To overcome this inconsistency, we apply lemma~\ref{lem:den_sem_change_var} about changing substituted fresh variable. When we do this, we lose the equality of values of representing functions on the first non-allocated variable. It is exactly the place where the proof of the strong version of the completeness theorem (without the restriction of domains of representing functions on allocated variables) fails (as explained in the section~\ref{equivalence}, the strong version of the completeness theorem does not hold).
\item In case of relational invocation we simply use the induction hypothesis with fewer unfoldings.
\end{itemize}

\section{\textsc{Coq} Specification}
\label{appendix_coq}

The \textsc{Coq} specification consists of three parts: specification of preliminary notions that we use in the semantics, specification of the language and semantics for interleaving search, specification of the language extended with cut and semantics for SLD resolution with cut (which repeats the previous part with only a few modifications).

Preliminary notions are notions from unification theory and definition of streams.

From unification theory we need (finitary) terms, substitutions, and operations on them (application to a term, composition). The most important notion we have to formalize here is the most general unifier. As it was described in the section~\ref{specification}, we define it as an inductive relation.

\begin{lstlisting}[language=Coq]
  Inductive mgu : term -> term -> option subst -> Set := ...
\end{lstlisting}

It uses an auxiliary function that finds the leftmost mismatch of the two terms (if there is any).

Then we prove by well-founded induction that this relation is a function and it satisfies the defining properties of MGU:

\begin{lstlisting}[language=Coq]
  Lemma mgu_result_$\texttt{exists}$ : forall t1 t2, {r & mgu t1 t2 r}.
  Lemma mgu_result_unique : forall t1 t2 r r',
      mgu t1 t2 r -> mgu t1 t2 r' ->r = r'.
  Definition unifier (s : subst) (t1 t2 : term) : Prop :=
    apply_subst s t1 = apply_subst s t2.
  Lemma mgu_unifies: forall t1 t2 s,
    mgu t1 t2 (Some s) -> unifier s t1 t2.
  Definition more_general (m s : subst) : Prop :=
    exists (s' : subst),
    forall (t : term),
      apply_subst s t = apply_subst s' (apply_subst m t).
  Lemma mgu_most_general : forall t1 t2 m,
      mgu t1 t2 (Some m) ->
      forall (s : subst), unifier s t1 t2 -> more_general m s.
  Lemma mgu_non_unifiable : forall t1 t2,
      mgu t1 t2 None -> forall s,  ~ (unifier s t1 t2).
\end{lstlisting}

For this well-founded induction we use the number of free variables in argument terms as a well-founded order on pairs of terms:

\begin{lstlisting}[language=Coq]
  Definition terms := term * term.
  Definition fvOrder (t : terms) :=
    length (union (fv_term (fst t)) (fv_term (snd t))).
  Definition fvOrderRel (t p : terms) :=
    fvOrder t < fvOrder p.
  Lemma fvOrder_wf : well_founded fvOrderRel.
\end{lstlisting}

Possibly infinite streams is a coinductive data type:

\begin{lstlisting}[language=Coq]
  Context {A : Set}.
  CoInductive stream : Set :=
  | Nil : stream
  | Cons : A -> stream -> stream.
\end{lstlisting}

However, some of its properties we are working with make sense only when defined inductively:

\begin{lstlisting}[language=Coq]
  Inductive in_stream : A -> stream -> Prop := ...
  Inductive finite : stream -> Prop := ...
\end{lstlisting}

For the equality of streams we need to define a new coinductive proposition instead of using the standard syntactic equality in order for coinductive proofs to work~\cite{CPDT}:

\begin{lstlisting}[language=Coq]
  CoInductive equal_streams : stream -> stream -> Prop :=
  | eqsNil  :  equal_streams Nil Nil
  | eqsCons : forall h1 h2 t1 t2,
                       h1 = h2 ->
                       equal_streams t1 t2 ->
                       equal_streams (Cons h1 t1) (Cons h2 t2).
\end{lstlisting}

Here we also define a relation for one-by-one interleaving of streams which we will use to describe the evaluation of dijunstions in the operational semantics:

\begin{lstlisting}[language=Coq]
  CoInductive interleave : stream -> stream -> stream -> Prop :=
  | interNil : forall s s', equal_streams s s' -> interleave Nil s s'
  | interCons : forall h t s rs,
                interleave s t rs ->
                interleave (Cons h t) s (Cons h rs).
\end{lstlisting}

This allows us to prove the expected properties of interleaving in a more general setting of arbitrary streams.

\begin{lstlisting}[language=Coq]
  Lemma interleave_in : forall s1 s2 s,
    interleave s1 s2 s ->
    forall x, in_stream x s <-> in_stream x s1 \/ in_stream x s2.
  Lemma interleave_finite : forall s1 s2 s,
    interleave s1 s2 s ->
    (finite s <-> finite s1 /\ finite s2).
\end{lstlisting}

The syntax of the language can be formalized in \textsc{Coq} in a natural way via inductive data types. Thanks to the higher-order syntax we need to work explicitly only with semantic variables. We use type ``\lstinline[language=Coq]{name}'' for all named entities (variables, constructors, and relations), and we define it simply as natural numbers.

As it was noted in the section~\ref{specification} all terms in our specification have arities either zero or two:

\begin{lstlisting}[language=Coq] 
   Inductive term : Set :=
   | Var : name -> term
   | Cst : name -> term
   | Con : name -> term -> term -> term.
\end{lstlisting}

And all relations have exactly one argument:

\begin{lstlisting}[language=Coq]
   Definition rel : Set := term -> goal.
\end{lstlisting}

We introduce one additional auxilary type of goals~--- \emph{failure}~--- for deliberately unsuccessful computation (empty stream). As a result, the definition of goals looks as follows:

\begin{lstlisting}[language=Coq] 
   Inductive goal : Set :=
   | Fail   : goal
   | Unify  : term -> term -> goal
   | Disj   : goal -> goal -> goal
   | Conj   : goal -> goal -> goal
   | Fresh  : (name -> goal) -> goal
   | Invoke : name -> term -> goal.
\end{lstlisting}

Note the use of HOAS for fresh variable introduction.

We define sets of free variables for terms naturally as \textsc{Coq} sets:

\begin{lstlisting}[language=Coq,mathescape=true] 
   Fixpoint fv_term (t : term) : $\mbox{\small{set}}$ name :=
     ...
\end{lstlisting}

And we use them to define ground terms as a subset type:

\begin{lstlisting}[language=Coq]
   Definition ground_term : Set :=
     {t : term | fv_term t = empty_set name}.
\end{lstlisting}

However, for goals it was more convenient to define a set of free variables as a proposition:

\begin{lstlisting}[language=Coq]
   Inductive is_fv_of_goal (n : name) : goal -> Prop :=
     ...
\end{lstlisting}

We use it to state our two explicit requirements on the syntax representation: the consistency of bindings and the absence of unbound variables in the definitions of relations. The second one is easy to describe:

\begin{lstlisting}[language=Coq] 
  Definition closed_goal_in_context 
    (c : list name) (g : goal) : Prop :=
      forall n, is_fv_of_goal n g -> In n c.
  Definition closed_rel (r : rel) : Prop :=
    forall (arg : term),
    closed_goal_in_context (fv_term arg) (r arg).
\end{lstlisting}

The consistency is based on variable renaming. It turned out to be rather non-trivial to define regular variable renaming for goals in higher-order syntax, but for our purposes
a weaker version, which deals only with non-free variables, is sufficient:

\begin{lstlisting}[language=Coq]
   Inductive renaming (old_x : name) (new_x : name) :
       goal -> goal -> Prop :=
   ...
   | rFreshNFV : forall fg,
                 (~is_fv_of_goal old_x (Fresh fg)) ->
                 renaming old_x new_x (Fresh fg) (Fresh fg)
   | rFreshFV : forall fg rfg,
                (is_fv_of_goal old_x (Fresh fg)) ->
                (forall y, (~is_fv_of_goal y (Fresh fg)) ->
                      renaming old_x new_x (fg y) (rfg y)) ->
                renaming old_x new_x (Fresh fg) (Fresh rfg)
   ...
\end{lstlisting}

The consistency for goals and relations then can be defined as follows:

\begin{lstlisting}[language=Coq]
   Definition consistent_binding (b : name -> goal) : Prop :=
     forall x y, (~ is_fv_of_goal x (Fresh b)) ->
           renaming x y (b x) (b y).
   Inductive consistent_goal : goal -> Prop :=
     ...
   Definition consistent_function
     (f : term -> goal) : Prop :=
     forall a1 a2 t, renaming a1 a2 (f t)
                     (f (apply_subst [(a1, Var a2)] t)).
   Definition consistent_rel (r : rel) : Prop :=
     forall (arg : term), consistent_goal (r arg) /\
                     consistent_function r.
\end{lstlisting}

In the snippet above the ``\lstinline[language=Coq]{consistent_goal}'' property inductively ensures that all bindings occurring
in the goal are consistent and ``\lstinline[language=Coq]{apply_subst [(a1, Var a2)] t}'' in ``\lstinline[language=Coq]{consistent_function}''
definition renames a variable \lstinline[language=Coq]{a1} into in \lstinline[language=Coq]{a2} term \lstinline[language=Coq]{t}.

We set an arbitrary environment (a map from a relational symbol to a definition of relation with described requirements) to use further throughout the formalization.
Failure goals allow us to define it as a total function:

\begin{lstlisting}[language=Coq]
   Definition def : Set := 
     {r : rel | closed_rel r /\ consistent_rel r}.
   Definition env : Set := name -> def.
   Axiom Prog : env.
\end{lstlisting}

To formalize denotational semantics in \textsc{Coq} we can define representing functions simply as \textsc{Coq} functions:

\begin{lstlisting}[language=Coq]
   Definition repr_fun : Set := name -> ground_term.
\end{lstlisting}

We define the semantics via the inductive proposition ``\lstinline|in_denotational_sem_goal|'' (with notation ``\lstinline[mathescape=true]{[| $\bullet$ , $\bullet$ |]}'')
such that

\[
\forall g,\mathfrak{f}\;:\;\mbox{\lstinline|in_denotational_sem_goal|}\;g\;\mathfrak{f}\Longleftrightarrow\mathfrak{f}\in\sembr{g}
\]

The head of the definition is as follows:

\begin{lstlisting}[language=Coq,morekeywords={where,at,level}]
  Reserved Notation "[| g , f |]" (at level 0).
  Inductive in_denotational_sem_goal :
    goal -> repr_fun -> Prop :=
    ...
  where "[| g , f |]" := (in_denotational_sem_goal g f).
\end{lstlisting}

and the body just goes through the cases shown in Fig.~\ref{denotational_semantics_of_goals}.

We also need to explicitly define the step-indexed version of denotational semantics described in the section~\ref{equivalence}:

\begin{lstlisting}[language=Coq,morekeywords={where,at,level}]
  Reserved Notation "[| n | g , f |]" (at level 0).
  Inductive in_denotational_sem_lev_goal :
    nat -> goal -> repr_fun -> Prop :=
    ...
  | dslgInvoke : forall l r t f,
      [| l  | proj1_sig (Prog r) t , f |] ->
      [| S l | Invoke r t , f |]
  where "[| n | g , f |]" :=
    (in_denotational_sem_lev_goal n g f).
\end{lstlisting}

Recall that the environment ``\lstinline[language=Coq]|Prog|'' maps every relational symbol to the definition of relation,
which is a pair of a function from terms to goals and a proof that it is closed and consistent.
So ``\lstinline[language=Coq]|(proj1_sig (Prog r) t)|'' here simply takes the body of the corresponding relation.

The lemma relating step-indexed and conventional denotational semantics in \textsc{Coq} looks as follows:

\begin{lstlisting}[language=Coq] 
  Lemma in_denotational_sem_some_lev:
    forall (g : goal) (f : repr_fun),
      [| g , f |] -> exists l, [| l | g , f |].
\end{lstlisting}

We prove two important properties of the denotational semantics: lemma~\ref{lem:den_sem_change_var} that states that the choice of a subsituted fresh variable doesn't matter, and the closedness condition (lemma~\ref{lem:closedness_condition}). The second proof is by a straightforward structural induction, the structure of the first one is described in the section~\ref{appendix_den_sem_change_var_proof}.

For operational semantics, the described transition relation can be encoded naturally as an inductively defined proposition (here ``\lstinline|nt_state|''
stands for a non-terminal state and ``\lstinline|state|''~--- for arbitrary one):

\begin{lstlisting}[language=Coq]
  Inductive eval_step :
    nt_state -> label -> state -> Set := ...
\end{lstlisting}

We state the fact that our system is deterministic through the existence and uniqueness of a transition for every state:

\begin{lstlisting}[language=Coq]
  Lemma eval_step_$\texttt{exists}$ : forall (nst : nt_state),
    {l : label & {st : state & eval_step nst l st}}.
  Lemma eval_step_unique : forall (nst : nt_state),
      (l1 l2 : label)
      (st1 st2 : state'),
    eval_step nst l1 st1 ->
    eval_step nst l2 st2 ->
    l1 = l2 /\ st1 = st2.
\end{lstlisting}

Then we define a trace coinductively as a stream of labels in transition steps and prove that there exists a unique trace from any extended state:

\begin{lstlisting}[language=Coq]
  Definition trace : Set := $@$stream label.
  CoInductive op_sem : state -> trace -> Set :=
  | osStop : op_sem Stop Nil
  | osState : forall nst l st t,
      eval_step nst l st ->
      op_sem st t ->
      op_sem (State nst) (Cons l t).
  Lemma op_sem_$\texttt{exists}$ (st : state):
    {t : trace & op_sem st t}.
  Lemma op_sem_unique:
    forall st t1 t2,
      op_sem st t1 ->
      op_sem st t2 ->
      equal_streams t1 t2.
\end{lstlisting}

And we prove the important properties of operational semantics (lemmas~\ref{lem:well_formedness_preservation}, \ref{lem:sum_answers}, \ref{lem:prod_answers} and \ref{lem:disj_termination}).

Finally, we prove both soundness and completeness of the operational semantics of the interleaving search w.r.t. the denotational one.
The statements of the theorems are as follows:

\newpage \begin{lstlisting}[language=Coq]]
  Theorem search_correctness:
    forall (g : goal) (k : nat) (f : repr_fun) (t : trace),
      closed_goal_in_context (first_nats k) g ->
      op_sem (State (Leaf g empty_subst k)) t ->
      {| t , f |} ->
      [| g , f |].
  Theorem search_completeness:
    forall (g : goal) (k : nat) (f : repr_fun) (t : trace),
      consistent_goal g ->
      closed_goal_in_context (first_nats k) g ->
      op_sem (State (Leaf g empty_subst k)) t ->
      [| g , f |] ->
      exists (f' : repr_fun),
        {| t , f' |} /\
        forall (x : var), In x (first_nats k) -> f x = f' x.
\end{lstlisting}

Note that we need the consistency requirement only for completeness, not for soundness.

The proof of soundness is by a straightforward structural induction, it is divided between lemmas~\ref{lem:answer_soundness} and \ref{lem:next_state_soundness}. The proof of completeness is described in the section~\ref{appendix_gen_completeness_proof}.

The specification of semantics for SLD resolution with cut simply repeats the just described specification of semantics for interleaving search with minimal modification. Specifically, we introduce one additional type of goal (\lstinline|Cut|), in denotational semantics cuts correspond to the whole domain of representing functions, in operational semantics we distinguish two kinds of \lstinline|Sum| state, introduce possible cut signal into the definition of transitions and add rules described in the section~\ref{appendix_cut} to the transition system. 

\begin{lstlisting}
  Inductive cutting_mark : Set :=
  | StopCutting : cutting_mark
  | KeepCutting : cutting_mark.
  Inductive nt_state : Set :=
  ...
  | Sum  : cutting_mark -> nt_state -> nt_state -> nt_state
  ...
  Inductive cut_signal : Set :=
  | NoCutting  : cut_signal
  | YesCutting : cut_signal.
  Inductive eval_step_SLD :
     nt_state -> label -> cut_signal -> state -> Set := ...
\end{lstlisting}

We then repeat the soundness proof for these semantics with minor changes (and skip the completeness proof, since it does not hold for SLD resolution).

\section{SLD Resolution with Cut}
\label{appendix_cut}

\begin{figure*}[t]
\[
  \begin{array}{cr}
    \dfrac{s_1 \xrightarrow{\step} \Diamond}{(s_1 \circledast s_2) \xrightarrow{\step} s_2} & \ruleno{AstStop}\\
    \dfrac{s_1 \xrightarrow{r} \Diamond}{(s_1 \circledast s_2) \xrightarrow{r} s_2} & \ruleno{AstStopAns}\\
    \dfrac{s_1 \xrightarrow{\step} s'_1}{(s_1 \circledast s_2) \xrightarrow{\step} (s_2 \circledast s'_1)} &\ruleno{AstStep}\\
    \dfrac{s_1 \xrightarrow{r} s'_1}{(s_1 \circledast s_2) \xrightarrow{r} (s_2 \circledast s'_1)} &\ruleno{AstStepAns}\\
  \end{array}
\]
\caption{Rules for ``$\circledast$'' evaluation}
\label{asterisk-rules}
\end{figure*}

\begin{figure*}[t]
\[
\begin{array}{crc|ccr}
  \dfrac{s_1 \xrightarrow{\step}_c \Diamond}{(s_1 \oplus s_2) \xrightarrow{\step} \Diamond} & \ruleno{SumStopC}& & &
  \dfrac{s_1 \xrightarrow{r}_c \Diamond}{(s_1 \oplus s_2) \xrightarrow{r} \Diamond} & \ruleno{SumStopAnsC}\\
  \dfrac{s_1 \xrightarrow{\step}_c s'_1}{(s_1 \oplus s_2) \xrightarrow{\step} s'_1} &\ruleno{SumStepC}& & &
  \dfrac{s_1 \xrightarrow{r}_c s'_1}{(s_1 \oplus s_2) \xrightarrow{r} s'_1} &\ruleno{SumStepAnsC}\\
  \dfrac{s_1 \xrightarrow{\step}_c \Diamond}{(s_1 \circledast s_2) \xrightarrow{\step}_c \Diamond} & \ruleno{AstStopC}& & &
  \dfrac{s_1 \xrightarrow{r}_c \Diamond}{(s_1 \circledast s_2) \xrightarrow{r}_c \Diamond} & \ruleno{AstStopAnsC}\\
  \dfrac{s_1 \xrightarrow{\step}_c s'_1}{(s_1 \circledast s_2) \xrightarrow{\step}_c s'_1} &\ruleno{AstStepC}& & &
  \dfrac{s_1 \xrightarrow{r}_c s'_1}{(s_1 \circledast s_2) \xrightarrow{r}_c s'_1} &\ruleno{AstStepAnsC}\\
  \dfrac{s \xrightarrow{\step}_c \Diamond}{(s \otimes g) \xrightarrow{\step}_c \Diamond} & \ruleno{ProdStopC}& & &
  \dfrac{s \xrightarrow{(\sigma, n)}_c \Diamond}{(s \otimes g) \xrightarrow{\step}_c \inbr{g, \sigma, n}} & \ruleno{ProdStopAnsC}\\
  \dfrac{s \xrightarrow{\step}_c s'}{(s \otimes g) \xrightarrow{\step}_c (s' \otimes g)} &\ruleno{ProdStepC}& & &
  \dfrac{s \xrightarrow{(\sigma, n)}_c s'}{(s \otimes g) \xrightarrow{\step}_c (\inbr{g, \sigma, n} \circledast (s' \otimes g))} &\ruleno{ProdStepAnsC}
\end{array}
\]
\caption{Cut signal propagation rules}
\label{cut-signal-propagation}
\end{figure*}

The semantics for SLD resolution with cut is built upon our usual semantics for SLD resolution described in section~\ref{sld} using a few additional constructs and rules for them.

We introduce one additional type of goal --- ``cut''. In denotational semantics, we interpret cut as success (thus, denotationally we treat all cuts as green).

Operationally, we modify SLD semantics in such a way that a cut cuts all other branches of all enclosing nodes, marked with ``$\oplus$'', up to
the moment when the evaluation of the disjunct, containing the cut, was started. It is easy to see that this node will always
be the nearest ``$\oplus$'', derived from the disjunction. Unfortunately, in the tree other ``$\oplus$'' nodes can
appear due to the evaluation of ``$\otimes$'' nodes, thus we need a way to distinguish these two sorts of ``$\oplus$''. We introduce a separate kind of state ``$\circledast$'' for the sums of goals created during the evaluation of ``$\otimes$'' nodes.

Therefore, in the semantics the rule \textsc{[ProdStepAns]} is replaced with the following.

\[
\begin{array}{cr}
  \dfrac{s \xrightarrow{(\sigma, n)} s'}{(s \otimes g) \xrightarrow{\circ} (\inbr{g, \sigma, n} \circledast (s' \otimes g))} & \ruleno{ProdStepAns} 
\end{array}
\]
\vskip3mm

The rules for ``$\circledast$'' evaluation mirror those for ``$\oplus$'' (see Fig.~\ref{asterisk-rules}).

To perform the cutting of the branches, we introduce a separate kind of transitions to propagate the signal for cutting (denoted by $\xrightarrow{l}_c$). The signal is risen when a ``cut'' construct is encountered.

\[
\begin{array}{cr}
  \inbr{!, \sigma, n} \xrightarrow{(\sigma, n)}_c \Diamond &\ruleno{Cut} 
\end{array}
\]
\vskip3mm

When the signal is being propagated through ``$\oplus$'' and ``$\circledast$'' nodes, their right branches are cut out, and for ``$\circledast$'' the
propagation continues; in the case of ``$\otimes$'' nodes the signal is simply propagated (see Fig.~\ref{cut-signal-propagation}).

\end{document}